# The Formation of Bilobate Comet Shapes through Sublimative Torques

Taylor K. Safrit[1], Jordan K. Steckloff[1,2,3], Amanda S. Bosh[1], David Nesvorny[4], Kevin Walsh[4], Ramon Brasser[5], David A. Minton[6]

[1]Massachussetts Institute of Technology, Cambridge, MA
[2]University of Texas at Austin, Austin, TX
[3]Planetary Science Institute, Tucson, AZ
[4]Southwest Research Institute, Boulder, CO
[5]Earth Life Science Institute, Tokyo Institute of Technology, Tokyo, Japan
[6]Purdue University, West Lafayette, IN

*Abstract*
**Recent spacecraft and radar observations have found that ~70 percent of short-period comet nuclei, mostly Jupiter-family comets (JFCs), have bilobate shapes (two masses connected by a narrow neck). This is in stark contrast to the shapes of asteroids of similar sizes, of which ~14% are bilobate. This suggests that a process or mechanism unique to comets is producing these shapes. Here we show that the bilobate shapes of JFC nuclei are a natural byproduct of sublimative activity during their dynamical migration from their trans-Neptunian reservoir, through the Centaur population, and into the Jupiter family. We model the torques resulting from volatile sublimation during this dynamical migration and find that they tend to spin up these nuclei to disruption. Once disrupted, the rubble pile-like material properties of comet nuclei (tensile strengths of ~1–10 Pa and internal friction angles of ~35º) cause them to reform as bilobate objects. We find that JFCs likely experienced rotational disruption events prior to entering the Jupiter family, which could explain the prevalence of bilobate shapes. These results suggest that the bilobate shapes of observed comets developed recently in their history (within the past ~1–10 Myr), rather than during solar system formation or collisions during planet migration and residency in the trans-Neptunian population**.

## Background

Centaurs are an unstable, transitional population of icy bodies that orbit the Sun in the region of the giant planets (between ~5 and 30 AU from the Sun) in dynamically unstable orbits. Centaurs originate as inwardly scattered trans-Neptunian objects (TNOs) and are ultimately ejected or dynamically evolve into Jupiter-family comets (JFCs; semimajor axes less than 5 AU) over ~1–10 Myr timescales (e.g. Tiscareno & Malhotra 2003). Thus, JFCs, Centaurs, and their TNO source population are dynamically linked, and sometimes collectively referred to as ecliptic comets (Duncan, Levison, and Dones 2004). Although no Centaur has yet been observed in high resolution, in situ spacecraft missions have investigated both the Centaur source population (TNOs) and end product population (JFCs), providing an indirect probe into the processes evolving ecliptic comets as they pass through the Centaur region. Although only 14% of nearly 200 radar-observed near-Earth asteroids have bilobate shapes (Taylor et al. 2012; Benner et al. 2015), ~70% of the seven nuclei of short period comets (e.g., JFCs) observed thus far with sufficient detail to obtain detailed shape data have such bilobate shapes (Figure 1) (Hirabayashi et al. 2016), suggesting that either material properties and/or physical processes unique to comets are favoring bilobate shape formation.

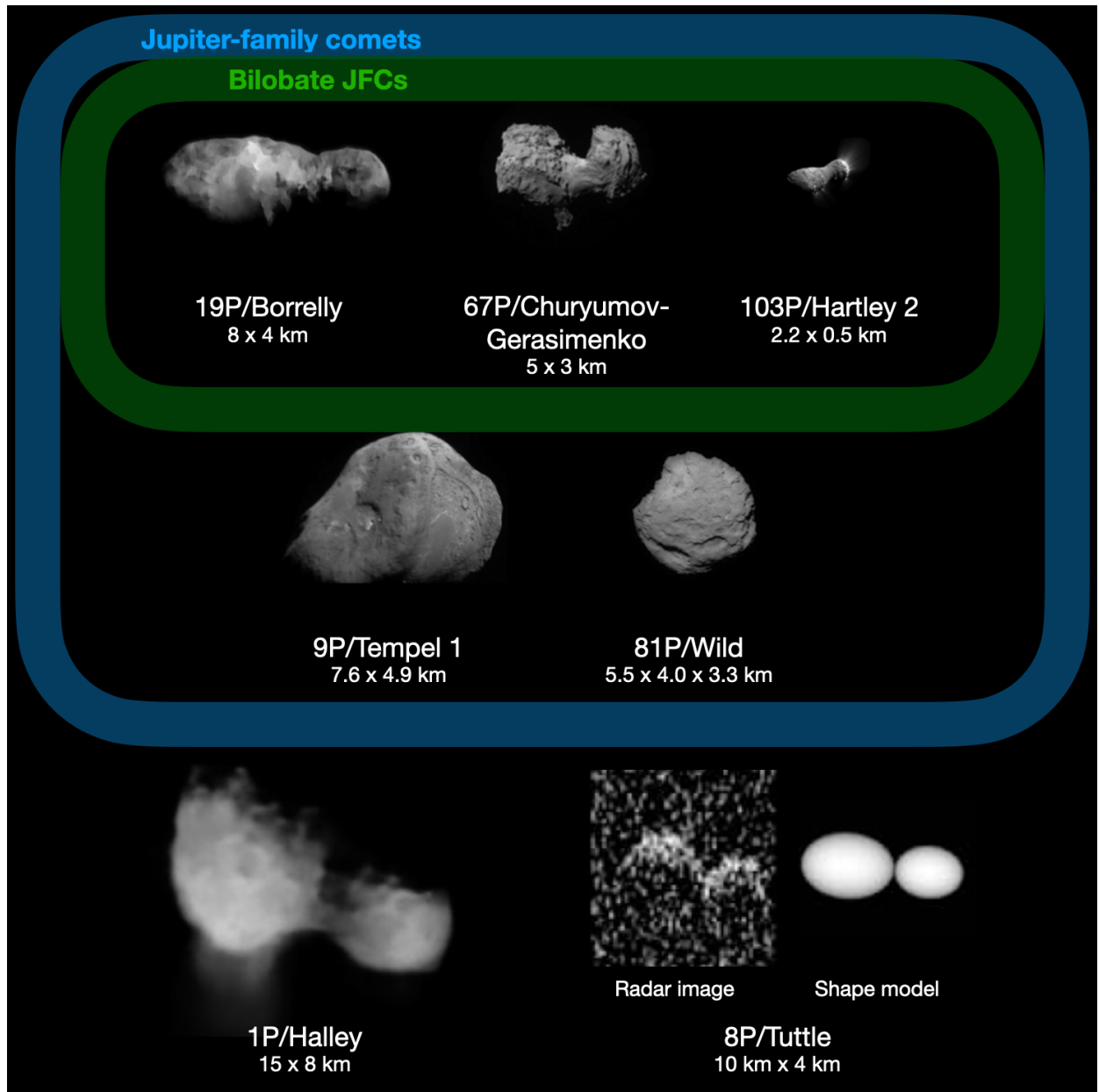

**Figure 1: Short-period comets with known shapes and their dimensions.**
Comets 8P/Tuttle and 1P/Halley are not classified as JFCs, but rather as Halley-family comets: highly thermally and dynamically evolved isotropic comets. 8P/Tuttle was observed as bilobate by the Arecibo radio telescope; the rest of these nuclei were observed by spacecraft. While detailed three-dimensional shape models do not exist for 19P/Borrelly and 1P/Halley, spacecraft observations suggest that they have bilobate shapes. Comets 9P/Tempel 1 and 81P/Wild are JFCs that are not observed to be bilobate (each has an oblate nucleus that is not considered elongated).

Recent work has shown that re-accretion following either catastrophic collisional (Schwartz et al. 2018) (Jutzi and Benz 2017) or rotational disruption (Sánchez and Scheeres 2016; 2018) can produce bilobate objects. The collisional pathway to bilobate shapes is well studied for low velocity (up to a few m/s) (Jutzi and Asphaug 2015), sub-catastrophic (~100s m/s) (Jutzi and Benz 2017), and catastrophic impacts (Schwartz et al. 2018). Remarkably, catastrophic collisions only result in minimal thermal processing of cometary volatiles (Schwartz et al. 2018), allowing impacts to produce these objects under the right conditions. Nevertheless, comets with diameters smaller than 4 km should experience a catastrophic impact (Morbidelli & Nesvorny 2020) with typical impact speeds of 2–4 km/s (Morbidelli & Rickman *2015*). However, such high impact speeds struggle to create the larger or more highly elongated bilobate comet nuclei observed by spacecraft (Schwartz et al. 2018), such as 19P/Borrelly or 103P/Hartley 2.

Alternatively, studies have found that the cold classical Kuiper Belt Object (KBO) (486958) Arrokoth obtained its shape through the collapse of a binary system, which merged to form one object (McKinnon et al. 2020). However, the orbital characteristics of the JFC reservoir population are markedly different from those of the cold classical KBOs, which have not experienced planetary encounters (Delsanti & Jewitt 2006). In contrast, the JFCs are thought to originate from the Scattered Disc population (Duncan & Levison 1997; Sarid et al. 2019) and the "stirred" parts of the classical Kuiper Belt (Sarid et al. 2019), which have experienced significant collisional evolution and encounters with the giant planets (Morbidelli & Nesvorny 2020) that can destroy such binary systems. As a result, the Scattered Disk population (the JFC reservoir population), is less likely to retain many similar binary systems, suggesting that some other, enigmatic process must be fundamentally altering the structure, shape, and surfaces of Jupiter-family comets.

In the Centaur region, ecliptic comets can begin to exhibit outgassing-driven cometary activity, such as outbursts (Trigo-Rodríguez et al. 2010; Rousselot et al. 2016), comae (Hartmann et al. 1990; Jewitt 2009), and fragment ejections (Rousselot 2008). Such activity can change the rotation state (Samarasinha and Mueller 2013; Steckloff and Jacobson 2016; Steckloff and Samarasinha 2018), shape (Hirabayashi et al. 2016), and surface morphology of cometary bodies (Vincent et al. 2017; Steckloff and Samarasinha 2018). Although too cold to be driven by water ice sublimation, Centaur activity can be driven by the sublimation of supervolatile species (such as CO and $CO_2$; Duffard 2002; Wierzchos, Womack, and Sarid 2017; cf. Jewitt 2009), or the crystallization of amorphous water ice (Jewitt 2009; Guilbert-Lepoutre 2012; Bauer et al. 2013). Asymmetries in outgassing generate torques (Samarasinha and Mueller 2013; Steckloff and Jacobson 2016) that can spin up nuclei to disruption.

Sánchez and Scheeres found that, after rotationally disrupting, small bodies with comet-like strength (Steckloff et al. 2015; Attree et al. 2018) and internal friction angles (Groussin et al. 2015; Steckloff and Samarasinha 2018) tend to reform into bilobate shapes (Sánchez and Scheeres 2016; 2018). Furthermore, once a comet becomes bilobate, further rotational spin up merely separates the lobes, which eventually reaccrete, preserving the bilobate shape (Hirabayashi et al. 2016). Finally, rotational disruption produces fragments with relative velocities small enough that the overwhelming majority of the disrupted body reaccretes into a single body (Sánchez and Scheeres 2016; 2018). This prevents rotational disruptions from producing significant numbers of fragments, consistent with observed JFC and TNO size-frequency distributions.

The different dependences of rotational inertia and sublimative torque on the size of the body makes rotational disruptions highly size-dependent (Steckloff and Jacobson 2016). Thus, smaller bodies spin up more rapidly, while larger bodies spin up more slowly. This leads to a critical radius, above which objects are simply too large to spin up to disruption over the dynamical lifetime of a Centaur.

Additionally, not all rotational disruptions necessarily produce bilobate shapes. Bilobate shape-formation requires that cohesion influence an object's internal stress state and disruption dynamics (Sánchez and Scheeres 2016; 2018). However, gravity dominates the cohesive forces of nuclei with radii greater than ~10 km (Sánchez and Scheeres 2014; Kokotanekova et al. 2017) favoring single masses rather than bilobate shapes. Nevertheless, the interior nucleus stress state does not scale analytically in the ~1-10 km transitional size regime (between cohesion-dominated and gravity-dominated dynamics; Sánchez and Scheeres 2020).

**Methods**

To understand the limits and constraints of sublimative torques on inducing Centaurs to disrupt (which may then reaccrete/deform into bilobate shapes), we studied how sublimative torques affect these objects during their dynamical migration into the JFC population. We first conducted numerical simulations of Centaur dynamics to obtain a representative sample of the dynamical pathways that JFCs take through the Centaur region. We then computed the sublimative torques experienced by Centaurs along these dynamical pathways, to understand the effects of rotational disruption statistically on the Centaur population as a whole.

*Dynamical Evolution Modeling*

Our dynamical modeling effort numerically integrated the evolution of orbits in the TNO population over the age of the Solar System, due to gravitational perturbations of the giant planets. This model neglects sublimative, non-gravitational forces acting on centaurs, which may additionally perturb these orbits. Rigorous testing has verified our TNO population model's accuracy against both the observed population of TNOs and reservoir formation theories (Nesvorný and Morbidelli 2012; Nesvorný, Vokrouhlický, and Roig 2016; Nesvorný et al. 2017). Our numerical integrator is the `swift_rmvs4` code from the Swift N-body integration package (Levison and Duncan 1994), that has been modified to account for the radial migration and damping of planetary orbits in the early Solar System via the exponential *e*-folding timescales (Nesvorný and Morbidelli 2012). Further details of our dynamical model, including a detailed description of the integration method, planet migration, initial orbital distribution of disk planetesimals, and comparison of results with the orbital structure of the TNO population can be found in Nesvorný et al. (2017).

Our numerical integrator tracks the orbital evolution of $10^6$ outer disk planetesimals from the onset of Neptune's migration to the present time. To improve statistics for modern day Centaurs, we produced 100 clones of objects that entered the Centaur region during the past 1 Gyr and repeated their dynamical evolution through the Centaur region. The cloning was accomplished by introducing a small (random) change of the velocity vector of a particle when it first evolves to an orbit with semimajor axis *a*<30 AU (i.e., enters the Centaur region). We recorded the cloned orbits every 100 years, until the object was either ejected from the Solar System or evolved into a JFC orbit. In this manner, we produced a total of ~$10^7$ Centaur orbits, resulting in a statistically significant sample of 346 representative dynamical pathways for JFCs through the Centaur region.

When the object is between 5 and 30 AU (i.e., in the Centaur region), we record the semimajor axis, eccentricity, and current heliocentric distance in 100-year intervals for each dynamical pathway (this interval is ~4–5 orders of magnitude shorter than the timescale of dynamical evolution through the Centaur region; e.g. di Sisto & Brunini 2007). This step unfortunately introduces an ambiguity into our simulations, as we do not record whether the particle is approaching aphelion or perihelion (or, equivalently, the mean, true, or eccentric anomalies). We therefore consider both cases for each dynamical pathway, which we ultimately find introduces a very small uncertainty into our constraints of rotational disruption.

*Sublimative Torque Modeling*
Sublimative torques arise from the asymmetric emission of sublimating volatiles from the irregular shapes of icy bodies. To compute their effects on the rotation state of Centaurs, we use the parameterized SYORP sublimative torque model (Steckloff and Jacobson 2016). The SYORP model uses the YORP effect of radiative torques as a template to compute the torques resulting from sublimation. Formulations of the YORP effect compute the magnitude of torques from a shape-dependent parameter and the magnitude of the radiation pressure at zero solar phase angle (Scheeres 2007; Rozitis and Green 2013). Similarly, the SYORP model uses first principles to solve for the mass flux and thermal velocity of sublimating volatiles leaving the surface, and thus the dynamic sublimative pressure that these gases exert on the surface. The SYORP model treats Centaurs as spheres of a single, pure volatile ice with negligibly low albedos (consistent with the observed low bond albedos of JFC nuclei), resulting in a pair of equations that can be solved simultaneously:

$$(1 - A) \frac{L_{solar}}{4\pi r_h^2 \lambda} cos\phi = \alpha_{(T)} \sqrt{\frac{m_{mol}}{2\pi RT}} P_{ref} e^{\frac{\lambda}{R}\left(\frac{1}{T_{ref}} - \frac{1}{T}\right)} \qquad (1)$$

$$P_{sub} = \frac{2}{3}(1 - A) \frac{L_{solar}}{4\pi r_h^2 \lambda} \left(\frac{8RT}{\pi m_{mol}}\right)^{\frac{1}{2}} cos\phi \qquad (2)$$

where $A$ is the Bond albedo of the material, $L_{solar}$ is the Sun's luminosity, $r_h$ is the heliocentric distance, $m_{mol}$ is the molar mass of the sublimating volatile, $R$ is the ideal gas constant, $\phi$ is the position of each sublimating area relative to the object's subsolar point, and $P_{ref}$ and $T_{ref}$ are a laboratory-determine pressure and temperature on the sublimative phase-change curve (Steckloff et al. 2015; Steckloff and Jacobson 2016). We chose CO and $CO_2$ as representative volatiles which undergo sublimation in the Centaur region. CO sublimation is active throughout the Centaur region, while $CO_2$ sublimation is active at heliocentric distances within approximately 13 AU. We chose an appropriate set of properties for each volatile used in our model, which are shown in Table 1 below. We also considered the effect of $H_2O$ sublimation but found that it is ultimately too refractory in the Centaur region to contribute to any significant changes in rotational states; we therefore neglect $H_2O$ from the remainder of our analyses.

| Volatile | $\lambda$ (J/mol) | $m_{mol}$ (kg/mol) | $P_{ref}$ (Pa) | $T_{ref}$ (K) |
|----------|-------------------|--------------------|-----------------|---------------|
| CO | 6720 | 0.028 | $3.45 \cdot 10^6$ | 132.572 |
| $CO_2$ | 27200 | 0.044 | $4.21 \cdot 10^{-2}$ | 102.5 |

**Table 1: Volatiles used and their properties.**
For our simulations, we used carbon monoxide and carbon dioxide as volatiles being sublimated from the surface of Centaurs. To first order, a molecule's volatility is determined by and inversely related to its latent heat of sublimation. Carbon monoxide is around 3 times more volatile than carbon dioxide by this metric; carbon dioxide is twice as volatile as water. These are the most relevant volatiles sublimating in the Centaur region. $\lambda$ is the volatile's latent heat of sublimation and $m_{mol}$ is its molar mass. $P_{ref}$ is an experimentally determined reference vapor pressure at reference temperature $T_{ref}$.

The SYORP model combines this sublimation pressure with a parameter $C_S$ (the "SYORP coefficient"), which encodes the effects of shape, spin pole orientation/obliquity, depth to sublimation fronts, and volatile-distribution on the magnitude of sublimative torques (Steckloff and Jacobson 2016; Steckloff and Samarasinha 2018). Thus, the angular accelerations resulting from sublimative torques are described by:

$$\frac{d\omega}{dt} = \frac{3 P_{sub} C_S}{4\pi \rho r^2} \tag{3}$$

where $\rho$ is the bulk density of the object (assumed to be uniform), $r$ is the effective radius of the object, and $C_S$ is the shape- and distribution-dependent SYORP coefficient. Although small, pristine icy bodies with surface-located sublimation fronts may have SYORP coefficients on the order of $\sim 10^{-4}$ to $10^{-3}$ (Steckloff and Jacobson 2016), more recent work computed the SYORP coefficients of JFC nuclei to lie in the range $\sim 10^{-5}$ to $10^{-4}$ (Steckloff et al. 2020). Nevertheless, we consider SYORP coefficients that span the entire range ($10^{-5}$ to $10^{-3}$) for completeness.

To compute the effects of sublimative torques during Centaur dynamical evolution, we applied this SYORP model to the computed Centaur dynamical pathways as they evolve inward. We used the Centaurs' orbital elements in each 100-year interval to compute the amount of time that it spent in 1 AU heliocentric distance bins in this range using the given orbital elements integrated forward for each 100-year timestep. For each body, we were able to compute a "best-case" (most spin-up) and "worst-case" (least spin-up) evolution given the uncertainty in orbital elements; we defined "best-case" as approaching perihelion at the start of the integration, and therefore spending more time near the Sun; "worst-case" was likewise defined as moving away from perihelion at the start of the integration and therefore spending less time near the Sun. In this way, we calculated 692 possible dynamical pathways from the 346 simulated pathways provided. Because sublimative changes in a Centaur's rotation state occur over secular (rather than orbital) timescales, this method is equivalent to computing the angular acceleration resulting from sublimative torques (in 1 AU intervals) and integrating over an entire orbit to compute the change in angular speed over a single orbit.

We calculated the total accrued angular velocity over an orbital evolution by summing the angular velocities accrued in each 1 AU heliocentric distance bin (given by multiplying the time spent at that distance, with minimum $dt$ of 100 years, by the angular acceleration at that distance):

$$\omega_{final} = \Sigma \frac{d\omega}{dt} dt \qquad (4)$$

We assumed that rotation state changes are solely the result of sublimative torques, ignoring the possibility of collisions or any other mechanism, and that such torques monotonically increased the angular velocity of each body from a non-rotating starting state. We also assumed that spin state changes were cumulative over the dynamical evolution of a Centaur, and therefore we were interested only in time spent at each heliocentric distance. With this method, we calculated the total change in rotation state (angular speed) for each of the 692 orbital cases as a function of object radius, as well as for a computed average orbital evolution (found by taking the average of the time spent in each 1 AU heliocentric distance bin for each of our orbits). The accrued angular velocity for this average orbit is shown as a function of object radius in Figure 1.

We compared this angular velocity to critical angular velocities for breakup in both strength- and gravity-dominated regimes. In the strength-dominated regime, the critical angular velocity $\omega_{crit}$ is given by:

$$\omega_{crit} = \sqrt{\frac{2\sigma_t}{\rho r^2}} \qquad (5)$$

where $r$ is the radius of the object, $\sigma_t$ is the tensile material strength and $\rho$ is the material density. For cometary nuclei, we assumed a tensile strength on the order of a few Pa, consistent with a rubble pile bound by Van der Waals forces (Sánchez and Scheeres 2014). In the gravity-dominated regime, the critical angular velocity $\omega_{crit}$ is given by:

$$\omega_{crit} = \sqrt{\frac{4\pi\rho G}{3}} \qquad (6)$$

where $G$ is the universal gravitational constant (Pravec and Harris 2000). Note that, unlike the strength-dominated critical angular velocity, the gravity-dominated critical velocity does not depend on the size of the object. Based on these critical angular velocities, we can compute a critical radius along every dynamical pathway for each combination of sublimating volatile and SYORP coefficient $C_S$. Below this critical radius, an object will disrupt during its migration. We used our model to compute the critical radius of disruption for each orbital pathway in our evolution set, and then computed a probability of disruption and critical size distribution (Figure 3) as a function of size, sublimating volatile, bulk density, and $C_S$.

**Results**

We find that sublimation-driven spin-up is capable of disrupting typical JFCs into bilobate shapes during their migration through the Centaur region. We fitted the distribution of critical sizes to Gaussian distributions, to obtain a statistical description of sizes below which objects rotationally disrupt during transit through the Centaur region (see Figure 4 for results for bulk densities of $\rho =$ 500 kg/m$^3$). Mean critical radii based on the fitted Gaussian for each histogram range from 7.2 km for $CO_2$ with $C_S$=10$^{-5}$ to 134.9 km for CO with $C_S$=10$^{-3}$. As expected, the high relative volatility of CO results in larger critical radii. Large critical size outliers appear to skew the distribution a bit. Nevertheless, we find that the distribution of critical sizes is typically much larger than the effective radii of JFCs with known shapes.

We also used our critical size computations to compute the probability of disruption in the Centaur region for each combination of parameters as a function of effective nucleus size, and found that objects with sizes typical for JFCs (on the order of ~1 km in radius) have high probabilities of disrupting in the Centaur region (Figure 3). Even for the weakest combination of parameters ($CO_2$-driven activity, $C_S = 10^{-5}$, high density of $\rho = 700$ kg/m$^3$), the probability of disruption is above 75% for all JFC nuclei with known shapes. For more potent combinations of parameters (CO-driven activity, $C_S = 10^{-4}$, low density of $\rho = 300$ kg/m$^3$), this rises to above 95% for such nuclei. Since such spin up tends to cause objects with JFC-like strength and internal friction angles to disrupt/deform into bilobate shapes (Sánchez and Scheeres 2016), these results suggest that rotational disruption of nuclei in the Centaur region may explain the prevalence of bilobate JFCs.

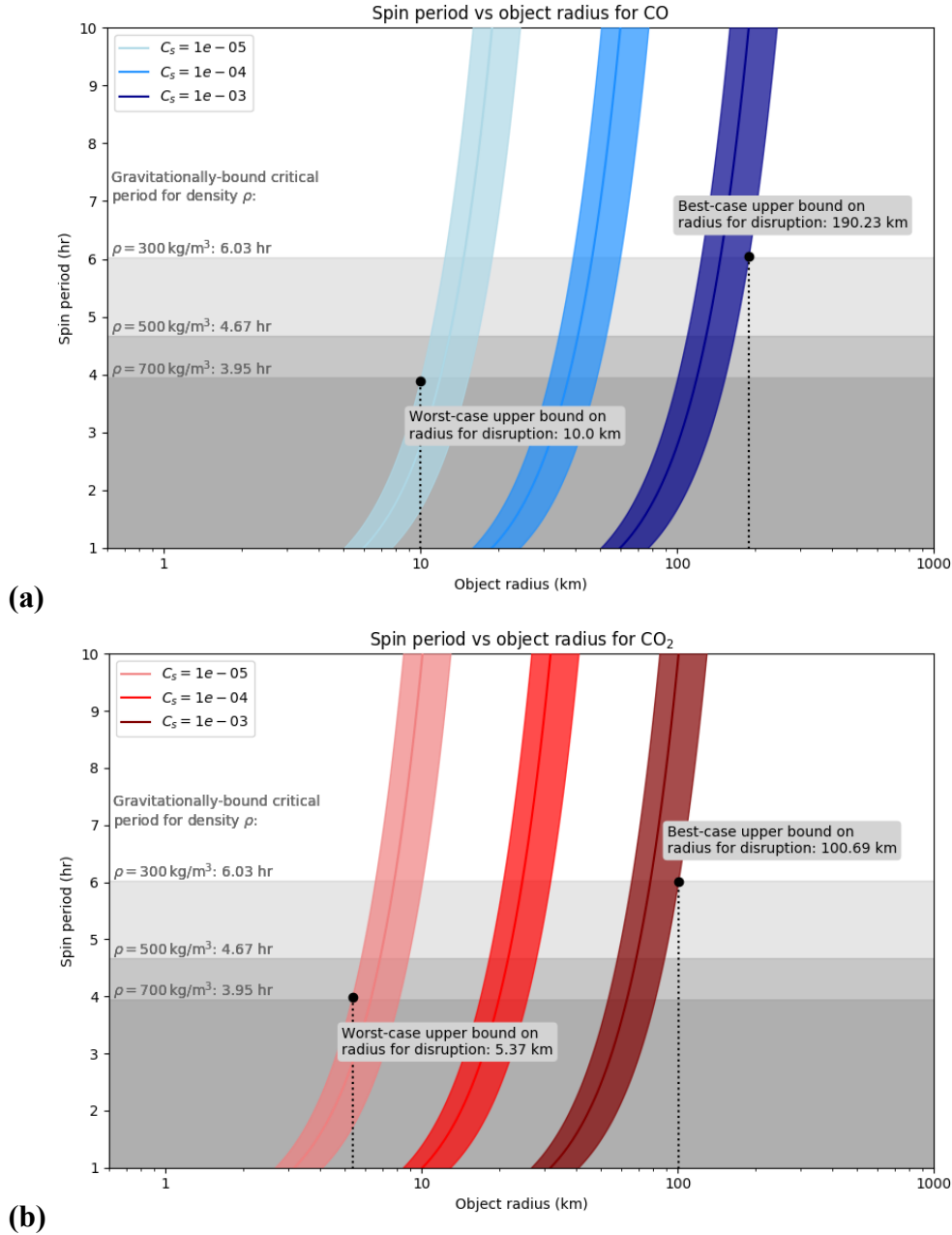

**(a)**

**(b)**

**Figure 2: Final spin period as a function of object radius for an average evolution.**
For CO, we find a critical radius between 10 km (for $C_S = 10^{-5}$ and body bulk density 700 kg/m³) and 190.2 km (for $C_S = 10^{-3}$ and body bulk density 300 kg/m³). For CO, we find a critical radius between 5.4 km (for $C_S = 10^{-5}$ and body bulk density 700 kg/m³) and 100.7 km (for $C_S = 10^{-3}$ and body bulk density 300 kg/m³). Each area corresponds to the upper limit on radius that can be disrupted for a range of densities and the indicated $C_S$: the low edge corresponds to a body bulk density of 700 kg/m³, the line through the area corresponds to a body bulk density of 500 kg/m³, and the upper edge corresponds to a body bulk density of 300 kg/m³. Objects with radii smaller than these values may spin up during their dynamical evolution to angular velocities large enough to cause breakup; after breakup, conditions are favorable for such bodies to reform in bilobate shapes.

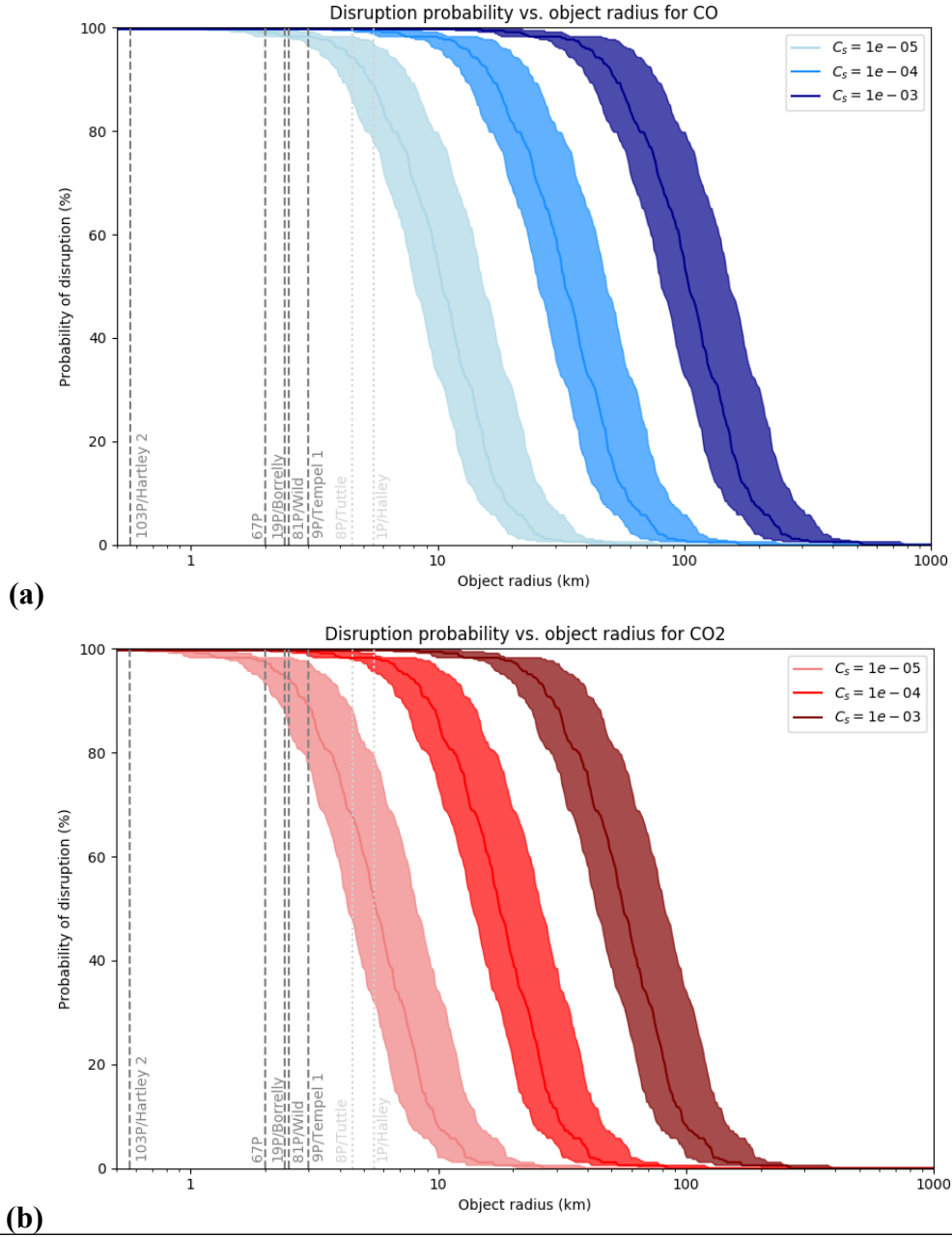

**(a)**

**(b)**

**Figure 3: Probability of disruption as a function of object radius for CO/CO2 sublimation.**
The probability of disruption is calculated statistically from our data set of Centaur orbits by comparing accrued angular velocity for each with the critical values as defined above and in Figure 1. We plot known effective radii of cometary nuclei on this result to show that our computed probabilities of disruption are high for even the weakest combinations of sublimating volatile and SYORP coefficient $C_S$. CO sublimates over the full Centaur region, while $CO_2$ sublimation is turned on within around 13 AU. Figure 2a shows probabilities for CO, while Figure 2b shows probabilities for $CO_2$. Each area corresponds to the probability of disruption as a function of radius for the indicated $C_S$ and a range of densities: the low edge corresponds to a body bulk density of 700 kg/m$^3$, the line through the area corresponds to a body bulk density of 500 kg/m$^3$, and the upper edge corresponds to a body bulk density of 300 kg/m$^3$.

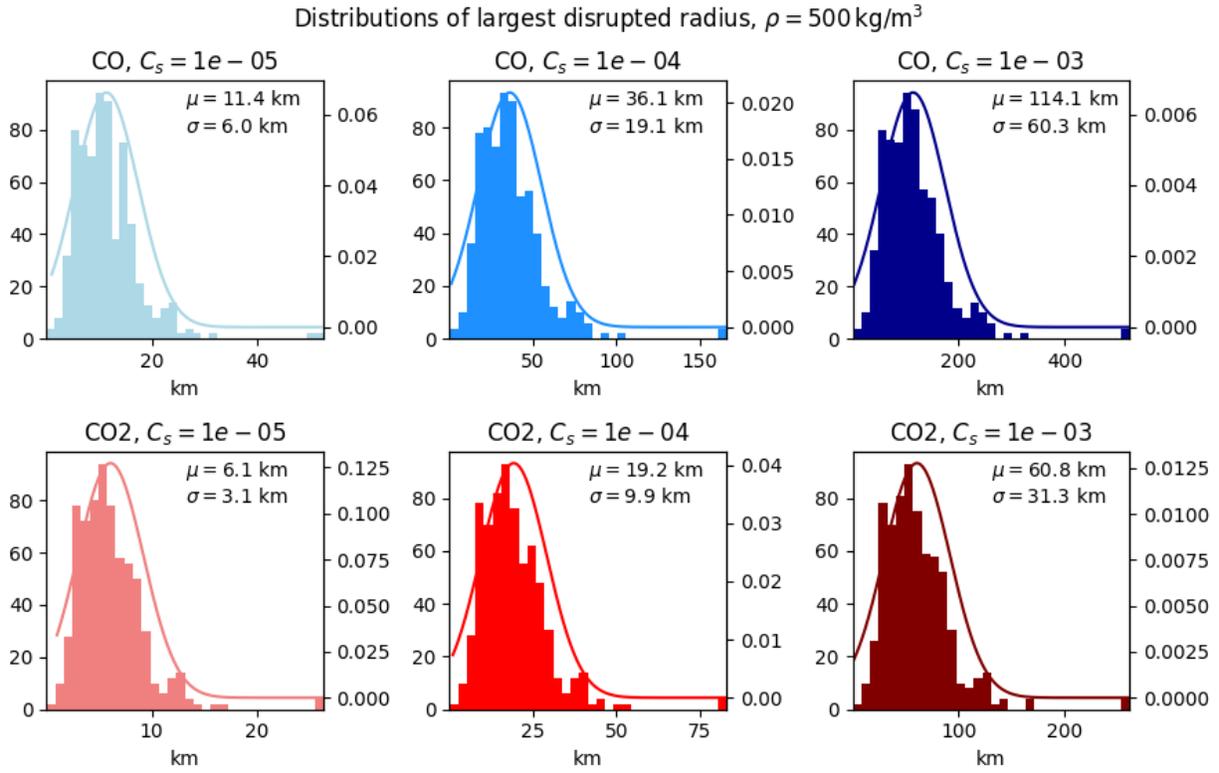

**Figure 4: Distribution of critical radius for disruption for each volatile/$C_S$ combination**
We plot histograms and fitted Gaussian distributions for each combination of volatile and SYORP coefficient $C_S$ and provide mean and standard deviation of each from the fitted Gaussian. Histogram bin sizes are computed with Freedman-Diaconis estimations. For each simulation, bulk density is set at 500 kg/m³ (the intermediate value of our defined density range as in Figures 1 and 2). Critical radius is defined as the largest body radius for which the accrued angular velocity over an evolution is sufficient to spin the body to disruption (assuming gravitationally bound bodies). Mean critical radii range from 6.1 km for $CO_2$ with $C_S = 10^{-5}$ to 114.1 km for CO with $C_S = 10^{-3}$. Each distribution is calculated from the same set of 692 orbital pathways, as described above; the distributions are therefore not independent but scaled corresponding to the combination of volatile and $C_S$.

**Discussion**

Our results suggest that the TNO reservoir of JFC progenitors may have a shape distribution that differs significantly from the JFCs. Small non-bilobate TNOs are likely to experience changes that reshape them into bilobate objects in the Centaur region. Nevertheless, were an SDO to already have a bilobate shape, such a shape would likely be preserved during rotational evolution (Hirabayashi et al. 2016), and would survive into the JFC population. Thus, we would only expect the JFC population to have significantly more bilobate objects than their trans-Neptunian reservoir if the population of small objects in this reservoir were not already dominated by bilobate shapes. Conversely, our results suggest that large objects are unlikely to change shape in the Centaur region. Thus, the shapes of the largest Centaurs are likely representative of the shape distribution of large objects in the JFC reservoir population. However, known JFCs have radii of a few km (Snodgrass et al. 2011); no centaurs or TNOs are known with such small sizes due to observational limitations. Therefore, the effect of this process, and the connection between the TNO, Centaur, and JFC size-frequency distributions is presently unknown.

Previous studies have found that close encounters with planets (such as with giant planets during migration through the Centaur region) could produce tidal forces that elongate objects, potentially into binaries or bilobate shapes (Richardson et al. 2002; Walsh 2018), or even disrupt them into fragments as happened with comet Shoemaker-Levy 9 (Asphaug and Benz 1996). Such dramatic tidal evolution requires close encounters on order the Roche radius or less (a few planetary radii), which happen but are extremely rare for any Centaur. Famously, such an encounter occurred between comet Shoemaker-Levy 9 and Jupiter in 1992 (Asphaug and Benz 1996). To explore the importance of this potential alternative mechanism to forming bilobate shapes in the Centaur region, we used the SyMBA symplectic *N*-body integrator (Duncan et al. 1998), modified to handle the typical close encounter distances between Centaurs and giant planets. We evolved the orbits of 144 known Centaurs for 10 Myr with a 0.5yr time step and found that Centaurs seldom encounter giant planets as distances less than ~100 planet radii (Figure 5). Thus, tidal disruptions contribute negligibly to the creation of bilobate comet nuclei.

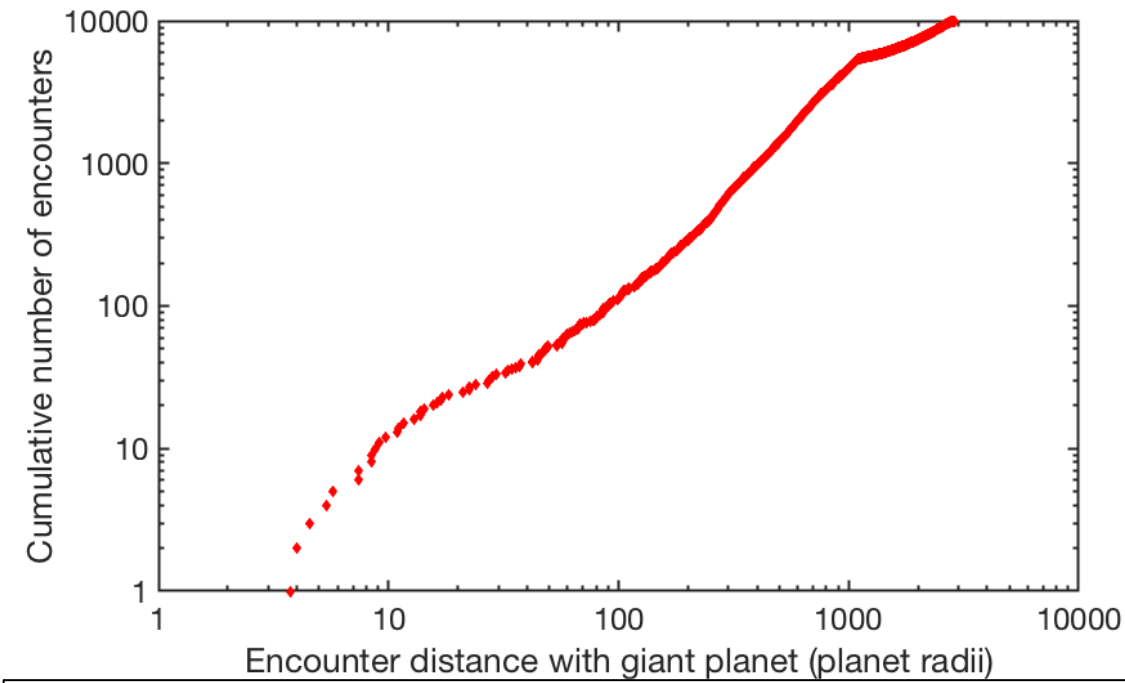

**Figure 5: Cumulative giant planet encounter distance frequency distribution.** We plot the close encounter distance of known centaurs with the giant planets over 10 Myr. Even encounters closer than ~100 planet radii, which is much greater than the encounter distances required to create bilobate shapes through tidal effects (a few planet radii), are rare.

Furthermore, our results suggest that any typical JFC has a *very* high likelihood of disrupting into a bilobate nucleus; this mechanism may even work *too* well, given that not all JFCs (such as 9P/Tempel 1 and 81P/Wild 2) are bilobate. This begs the question: why did these nuclei fail to become bilobate?

*9P/Tempel 1, 81P/Wild 2, and Non-bilobate Nuclei*

Rotational disruption produces bilobate shapes (rather than losses of surface materials) when cohesive forces are strong enough to induce structural failure in the interior of the body, rather than at the surface (Sánchez and Scheeres 2016). As objects increase in size, gravitational forces play an increasing role in binding the object together. By setting equations (5) and (6) equal to each other we can find the size at which nuclei transition from being predominantly bound by cohesion to predominantly bound by gravity:

$$r = \frac{1}{\rho} \sqrt{\frac{3\sigma_t}{2\pi G}} \tag{7}$$

For a tensile strength on the order of ~10 Pa and a density of ~500 kg/m$^3$, we obtain a radius of ~500 m. Above this radius, cohesion still helps to bind the nucleus, but plays a decreasing role as nuclei increase in radius. Above some size (perhaps an order of magnitude larger than this transitional size), gravitational forces sufficiently dominate cohesive forces to prevent cohesion from influencing the outcome of rotational disruption. 9P/Tempel 1 and 81P/Wild 2 may be in this category of objects small enough to rotationally disrupt, yet too large to reform/deform into a bilobate shape. Indeed, Tempel 1 seems to exhibit large scale layering throughout its nucleus (Thomas et al. 2007), which could be the result of its disruption.

To form our results, we assumed that all comets that spin up to disruption reform into bilobate shapes. However, previous numerical studies have shown that rotational spin-up of rubble piles does not necessarily produce bilobate shapes (Richardson et al. 2005; Richardson et al. 2009). These studies found that a rotationally disrupted aggregate's fate is sensitive to material cohesion: in strengthless cases, objects tend to elongate (Richardson et al. 2005), and in high-strength cases, objects tend to shed material (Richardson et al. 2009). However, in cases of low but non-zero cohesive strength (on the order of ~100 Pa), this work found that nucleus deformation can occur before failure, which could lead to the formation of binary or bilobate objects (Richardson et al. 2009). Sánchez and Scheeres (2016) found that bilobate formation also depends on internal friction angle and found that the cohesive strengths and internal friction angles that result in bilobate formation are ~0.6–600 Pa and ~25–35º, respectively. Nevertheless, it is plausible that some Centaurs could have material properties that lie outside of this range, which would preclude bilobate shape formation. Therefore, our results may overestimate the probability of bilobate formation.

Alternatively, 9P/Tempel 1 and 81P/Wild 2 could have become bilobate but lost their smaller lobes. Once a nucleus becomes bilobate, rotational spin up will cause the two lobes to separate (Hirabayashi et al. 2016). The fate of such lobes depends on the relative masses of the two lobes; if their mass ratio is greater than 0.2 (similar-size lobes), the two lobes remain gravitationally bound, and will reaccrete into a bilobate shape. However, if the lobes' mass ratio is less than 0.2, the lobes tend to be gravitationally unbound; the lobes separate permanently and the comet splits (Hirabayashi et al. 2016). If either 9P/Tempel 1 or 81P/Wild 2 had a sufficiently small lobe, it may have split from the larger lobe, leaving a single, non-bilobate lobe comprising the nucleus. However, the nucleus could then spin up again to bilobate formation; a process that could plausibly repeat until a stable bilobate nucleus with lobes comparable in size (mass ratio greater than 0.2) formed. Nuclei close to the critical disruption radius would be less likely to have sufficient time to experience multiple disruption cycles; if such nuclei only produced small lobes less than 20% the mass of the larger lobe and continued to escape, then the prevalence of bilobate shapes should generally decrease with increasing size near the disruption limit. Nevertheless, both of these comets are much smaller than the critical disruption limit, and previous modeling work tends to show that roughly equal lobes would be created (Sánchez and Scheeres 2016). We therefore consider this scenario to be less likely.

Assumptions made in our model may produce errors in the calculated sizes at which bilobate objects can form. For example, real bodies are finite in size and may be depleted in volatiles before they are able to disrupt. However, our model assumes that objects have an infinite supply of volatiles. To estimate this effect, we calculated the depth of ice that would be lost from a body undergoing an average Centaur evolution (as we used to create Figure 2), assuming a density of 500 kg/m$^3$. We found that such a typical Centaur would lose 340 km or 7100 km to $CO_2$ or CO sublimation, respectively, before entering the Jupiter family. No known Centaur is this large, and bodies with this amount of ice would not be able to spin up to disruption, suggesting that the actual size limits for bilobate formation are smaller than our calculated limits as given in Figure 4 (see the mean of each distribution).

We can estimate what these actual size limits are using a simple calculation. In our model, angular acceleration depends on $1/r^2$ (as in equation 3), so a body with half of the radius of another would require a fourth of the amount of ice to spin to disruption. Applying this same scaling to our modeled size limits, an object sublimating CO with SYORP coefficients $C_S = 10^{-4}$ and $C_S = 10^{-5}$ would need to have radius smaller than ~1 km and ~100 m, respectively, to be disrupted, while an object sublimating $CO_2$ would need to have radius smaller than ~200 m and ~20 m (respectively for the above $C_S$ values). This suggests that $CO_2$ is much more important than CO in spinning up JFCs (at the observed size range of ~1 km radius). Additionally, the distribution of bilobate JFCs is likely sensitive to the distribution of SYORP coefficients among Centaurs.

Errors and biases in estimates of $C_S$ values may also produce an overabundance of bilobate nuclei. The range of empirical $C_S$ values ($10^{-5}$–$10^{-4}$) was computed from observations of rotation state changes in JFC nuclei (Steckloff et al. 2020), either from lightcurve studies or spacecraft observations. However, such observations are inherently biased toward more active comets, which are more visible from Earth. However, the level and distribution of sublimative activity across the nucleus surface is folded into the $C_S$ parameter, with higher levels of activity favoring greater $C_S$ values, potentially biasing this sample toward larger $C_S$ values. In such a case, $C_S$ values for a representative population may plausibly be lower than $10^{-5}$–$10^{-4}$. As angular acceleration from sublimation scales linearly with $C_S$, a lower average $C_S$ would reduce the critical radius limit for disruption that we calculated. The probabilities of disruption for non-bilobate JFCs could in reality be sufficiently low to explain their existence.

Additionally, our $C_S$ values may overestimate the concentration of supervolatile species across the nucleus surface (the volatile abundance/distribution component of $C_S$). Presently, all empirically derived $C_S$ values are for $H_2O$ sublimation in the inner solar system, with the notable exception of 103P/Hartley 2, whose activity is dominated by $CO_2$ sublimation (Steckloff et al. 2020). Although volatile abundances within Centaurs are poorly constrained, CO and $CO_2$ are nevertheless typically significantly less abundant than $H_2O$ in cometary comae (Bockelée-Morvan et al. 2004), resulting in lower volatile production rates which drive sublimative torques. If we assume that these coma abundances reflect surface compositions, our model may overestimate CO or $CO_2$ production rates, which may be as much as an order of magnitude smaller than that of $H_2O$. This would ultimately result in smaller critical radii for disruption than computed. However, these reduced abundances would have a surprisingly small effect on the net sublimative torque experienced by comets. Most comet outgassing produces torques that stochastically cancel out; reducing volatile production rates produces a corresponding change in the measured net torque of only ~3–20% (Samarasinha & Mueller 2013; Mueller & Samarasinha 2018). Thus, this error is unlikely to produce significant errors in critical radii for disruption, which are already dominated by uncertainties in the SYORP parameter $C_S$.

Finally, our assumption of monotonic changes in spin rate may be overestimating actual spin rate changes. The analogous radiative YORP torques are thought to be stochastic or varying in magnitude and direction over time (Statler 2009; Cotto-Figueroa et al. 2015), causing such monotonic assumptions to overestimate maximum spin rates (specifically, angular speeds) by a small factor on the order of unity (Cotto-Figueroa et al. 2015). Sublimative torques likely follow a similar pattern, as rotation changes may activate new areas of sublimative activity (Steckloff & Samarasinha, 2018) that change the magnitude and direction of sublimative torques. These torque changes may be somewhat self-limiting due to stochastic torque cancelations (Samarasinha & Mueller, 2013). Nevertheless, if sublimative torque changes behave similarly to radiative torques, the critical sizes for each dynamical pathway (and thus the size of disrupted nuclei) may be overestimated by a small factor (on the order of unity). Given that 9P/Tempel 1 has an SYORP coefficient of only $1.22 \times 10^{-5}$ (Steckloff et al. 2020), such a shift could make it unlikely that Tempel 1 would have ever spun up to disruption. Similarly, comet 19P/Borrelly has a large SYORP coefficient (Steckloff et al. 2020); if 81P/Wild 2 has a smaller SYORP coefficient, then this time varying effect could explain both why Borrelly has a bilobate shape and Wild 2 does not.

*Split Comets*
Rotational disruption, perhaps driven by sublimative torques, has long been proposed as a mechanism for comet splitting (e.g. Boehnhardt 2000). For example, comet 3D/Biela split into two distinct objects in 1846 before disappearing, presumed disintegrated (Jenniskens and Vaubaillon 2007). If Biela was originally a bilobate comet with lobes having a mass ratio of less than 0.2, rotational disruption could have split Biela into a gravitationally unbound system, causing its lobes to separate into two visually distinct objects. A few such systems exist; recent observations have revealed another split comet candidate in comets 252P and BA14 (Li et al. 2017). More generally, Chen and Jewitt predict a splitting rate of cometary nuclei of 0.01 per comet per year (Chen and Jewitt 1994). Sublimation-driven rotational spin up may explain this rate of comet bursting; however, such an investigation is outside the scope of this work.

*Binary Comets*
Binary asteroids are thought to form through rotational disruption, in which an asteroid rotates rapidly enough for material to be rotationally shed from the equator before collecting into a small moon. The formation or rotational evolution of comet nuclei should similarly produce binary systems, if only temporarily. However, no comets or Centaurs are known to be binary systems[1]. One reason for this may be the differences in the strengths of the forces that evolve these systems. Small asteroid systems evolve primarily though radiative forces (e.g., YORP and binary-YORP or "BYORP"), which can cause the secondary to either spiral in or leave the system (Cuk and Burns 2005). A typical 1 km radius asteroid with a 0.2 km radius secondary ("moon") at 1AU would experience such evolution over a timescale of $\sim 10^4$–$10^5$ years (Steinberg and Sari 2011). However, sublimative forces are typically $\sim 10^5$ times stronger than radiative forces when sublimation is vigorous (Steckloff and Jacobson 2016), causing vigorously sublimating systems to evolve $\sim 10^3$–$10^4$ times faster, once accounting for the relative weakness of SYORP coefficients relative to their YORP coefficient analogues (Steckloff et al. 2020) and the relatively low density of cometary bodies relative to asteroids. Thus, sublimative forces would cause a comparable binary cometary system with a 1 km primary at 0.2 km secondary at 1 AU to evolve to either reaccrete into a binary system or separate over a $\sim 1$–100-year timescale. Such timescales are incredibly short, resulting in a significantly lower probability that a binary comet system would be observed relative to a binary asteroid system. Furthermore, the comae of vigorously sublimating objects could further obscure any binary system, rendering it unlikely that any binary cometary system would have ever been detected.

## Conclusions

Unlike the near-Earth asteroid population, the majority of short period comet nuclei have bilobate shapes, suggesting that material properties and/or physical processes unique to comets are favoring bilobate shape formation. Previously proposed mechanisms to explain this discrepancy involve either improbably low-velocity collisions between similarly sized objects, or high-velocity impacts that are inconsistent with the size-frequency distribution of the JFC reservoir population. We explore an alternative mechanism, in which sublimative torques rotationally disrupt JFC nuclei through the Centaur region from their trans-Neptunian reservoir population. Given the material properties of comet nuclei, such disruption events are likely to result in nuclei reforming or deforming into bilobate shapes. We find that this mechanism is extremely efficient at producing bilobate shapes among the population of JFCs with known sizes. Even our worst-case scenario produces a probability of disruption for the largest JFCs with known shapes of greater than 75%, with our best-case scenarios disrupting all JFCs with near certainty. Nevertheless, there is an unknown cutoff point in body size beyond which rotational disruption is unlikely to result in bilobate shapes; this size is presently unknown.

## Acknowledgments

We wish to thank the anonymous reviewers, whose comments greatly improved the quality and clarity of this work. J.K.S. was supported in part by NSF grant 1910275 and NASA award 80NSSC18K0497. D.N. was supported by NASA's Solar System Workings program. RB acknowledges financial assistance from JSPS Shingakujutsu Kobo (JP19H05071).

## References

---

[1] We do not consider split comets to be binary, as they are likely to be the gravitationally unbound result of cometary breakup.

Asphaug, Erik, and Willy Benz. 1996. "Size, Density, and Structure of Comet Shoemaker–Levy 9 Inferred from the Physics of Tidal Breakup." *Icarus* 121 (2): 225–48. https://doi.org/10.1006/icar.1996.0083.

Attree, N., O. Groussin, L. Jorda, D. Nébouy, N. Thomas, Y. Brouet, E. Kührt, et al. 2018. "Tensile Strength of 67P/Churyumov–Gerasimenko Nucleus Material from Overhangs." *Astronomy & Astrophysics* 611 (March): A33. https://doi.org/10.1051/0004-6361/201732155.

Bauer, James M., Tommy Grav, Erin Blauvelt, A. K. Mainzer, Joseph R. Masiero, Rachel Stevenson, Emily Kramer, et al. 2013. "Centaurs and Scattered Disk Objects in the Thermal Infrared: Analysis of WISE/ NEOWISE Observations." *The Astrophysical Journal* 773 (1): 22. https://doi.org/10.1088/0004-637X/773/1/22.

Benner, L. A. M., M. W. Busch, J. D. Giorgini, P. A. Taylor, and J.-L. Margot. 2015. "Radar Observations of Near-Earth and Main-Belt Asteroids." In *Asteroids IV*, edited by Patrick Michel, Francesca E. DeMeo, and William F. Bottke. University of Arizona Press. https://doi.org/10.2458/azu_uapress_9780816532131-ch009.

Bockelée-Morvan, D; Crovisier, J.; Mumma, M.J.; Weaver, H.A. 2004. "The Composition of Cometary Volatiles".  In *Comets II* Editors: Festou, M.; Keller, H.U.; Weaver, H.A.. University of Arizona Press - Tucson, AZ 391 - 423

Boehnhardt, Hermann. 2000. "Comet Splitting – Observations and Model Scenarios." *Earth, Moon, and Planets* 89 (1/4): 91–115. https://doi.org/10.1023/A:1021538201389.

Chen, J, and David Jewitt. 1994. "On the Rate at Which Comets Split." *Icarus* 108 (2): 265–71. https://doi.org/10.1006/icar.1994.1061.

Cotto-Figueroa, D.; Statler, T.S.; Richardson, D.C.; Tanga, P. (2015) Coupled Spin and Shape Evolution of Small Rubble-Pile Asteroids: Self-Limitation of the YORP Effect. *ApJ* 803:25 (18 pp.)

Cuk, M, and J Burns. 2005. "Effects of Thermal Radiation on the Dynamics of Binary NEAs." *Icarus* 176 (2): 418–31. https://doi.org/10.1016/j.icarus.2005.02.001.

Delsanti, A.; Jewitt, D. (2006) "The Solar System Beyond the Planets". In *Solar System Update*. Eds: P. Blondel and J. Mason

Duffard, R. 2002. "New Activity of Chiron: Results from 5 Years of Photometric Monitoring." *Icarus* 160 (1): 44–51. https://doi.org/10.1006/icar.2002.6938.

Duncan, M.J.; Levison, H.F. (1997) A disk of Sattered Icy Objects and the Origin of Jupiter-Family Comets. *Science* **276**, 1670 - 1672

Duncan, M.J.; Levison, H.F.; Lee, M.H. (1998) A Multiple Time step symplectic algorithm for integrating close encounters. *Astronomical Journal* 116, 2067 - 2077

Duncan, Martin, Harold Levison, and Luke Dones. 2004. "Dynamical Evolution of Ecliptic Comets." In *Comets II*. Space Science Series. University of Arizona Press.

Groussin, O., H. Sierks, C. Barbieri, P. Lamy, R. Rodrigo, D. Koschny, H. Rickman, et al. 2015. "Temporal Morphological Changes in the Imhotep Region of Comet 67P/Churyumov-Gerasimenko." *Astronomy & Astrophysics* 583 (November): A36. https://doi.org/10.1051/0004-6361/201527020.

Guilbert-Lepoutre, Aurélie. 2012. "Survival of Amorphous Water Ice on Centaurs." *The Astronomical Journal* 144 (4): 97. https://doi.org/10.1088/0004-6256/144/4/97.

Hartmann, William K., David J. Tholen, Karen J. Meech, and Dale P. Cruikshank. 1990. "2060 Chiron: Colorimetry and Cometary Behavior." *Icarus* 83 (1): 1–15. https://doi.org/10.1016/0019-1035(90)90002-Q.

Hirabayashi, Masatoshi, Daniel J. Scheeres, Steven R. Chesley, Simone Marchi, Jay W. McMahon, Jordan Steckloff, Stefano Mottola, Shantanu P. Naidu, and Timothy Bowling. 2016. "Fission and Reconfiguration of Bilobate Comets as Revealed by 67P/Churyumov–Gerasimenko." *Nature* 534 (7607): 352–55. https://doi.org/10.1038/nature17670.

Jenniskens, P., and J. Vaubaillon. 2007. "3D/Biela and the Andromedids: Fragmenting versus Sublimating Comets." *The Astronomical Journal* 134 (3): 1037–45. https://doi.org/10.1086/519074.

Jewitt, David. 2009. "The Active Centaurs." *The Astronomical Journal* 137 (5): 4296–4312. https://doi.org/10.1088/0004-6256/137/5/4296.

Jutzi, M., and E. Asphaug. 2015. "The Shape and Structure of Cometary Nuclei as a Result of Low-Velocity Accretion." *Science* 348 (6241): 1355–58. https://doi.org/10.1126/science.aaa4747.

Jutzi, M., and W. Benz. 2017. "Formation of Bi-Lobed Shapes by Sub-Catastrophic Collisions: A Late Origin of Comet 67P's Structure." *Astronomy & Astrophysics* 597 (January): A62. https://doi.org/10.1051/0004-6361/201628964.

Kokotanekova, R., C. Snodgrass, P. Lacerda, S. F. Green, S. C. Lowry, Y. R. Fernández, C. Tubiana, A. Fitzsimmons, and H. H. Hsieh. 2017. "Rotation of Cometary Nuclei: New Light Curves and an Update of the Ensemble Properties of Jupiter-Family Comets." *Monthly Notices of the Royal Astronomical Society* 471 (3): 2974–3007. https://doi.org/10.1093/mnras/stx1716.

Levison, Harold F., and Martin J. Duncan. 1994. "The Long-Term Dynamical Behavior of Short-Period Comets." *Icarus* 108 (1): 18–36. https://doi.org/10.1006/icar.1994.1039.

Li, Jian-Yang, Michael S. P. Kelley, Nalin H. Samarasinha, Davide Farnocchia, Max J. Mutchler, Yanqiong Ren, Xiaoping Lu, David J. Tholen, Tim Lister, and Marco Micheli. 2017. "The Unusual Apparition of Comet 252P/2000 G1 (LINEAR) and Comparison with Comet P/2016 BA $_{14}$ (PanSTARRS)." *The Astronomical Journal* 154 (4): 136. https://doi.org/10.3847/1538-3881/aa86ae.

Massironi, Matteo, Emanuele Simioni, Francesco Marzari, Gabriele Cremonese, Lorenza Giacomini, Maurizio Pajola, Laurent Jorda, et al. 2015. "Two Independent and Primitive Envelopes of the Bilobate Nucleus of Comet 67P." *Nature* 526 (7573): 402–5. https://doi.org/10.1038/nature15511.

McKinnon, W. B., D. C. Richardson, J. C. Marohnic, J. T. Keane, W. M. Grundy, D. P. Hamilton, D. Nesvorný, et al. 2020. "The Solar Nebula Origin of (486958) Arrokoth, a Primordial Contact Binary in the Kuiper Belt." *Science* 367 (6481): eaay6620. https://doi.org/10.1126/science.aay6620.

Morbidelli, A., and H. Rickman. 2015. "Comets as Collisional Fragments of a Primordial Planetesimal Disk." *Astronomy & Astrophysics* 583 (November): A43. https://doi.org/10.1051/0004-6361/201526116.

Morbidelli, A.; and D. Nesvorny. 2020. "Kuiper belt: Formation and evolution. The Trans-Neptunian Solar System", In *The Trans-Neptunian Solar System*. Ed: Prialnik, D.; Barucci, M.A.; Young, L. pp. 25-59

Nesvorný, David, and Alessandro Morbidelli. 2012. "Statistical Study of the Early Solar System's Instability with Four, Five, and Six Giant Planets." *The Astronomical Journal* 144 (4): 117. https://doi.org/10.1088/0004-6256/144/4/117.

Nesvorný, David, Joel Parker, and David Vokrouhlicky. 2018. "Bi-Lobed Shape of Comet 67P from a Collapsed Binary." *ArXiv Astro-Ph.EP*, April. arXiv:1804.08735v1.

Nesvorný, David, David Vokrouhlický, Luke Dones, Harold F. Levison, Nathan Kaib, and Alessandro Morbidelli. 2017. "Origin and Evolution of Short-Period Comets." *The Astrophysical Journal* 845 (1): 27. https://doi.org/10.3847/1538-4357/aa7cf6.

Nesvorný, David, David Vokrouhlický, and Fernando Roig. 2016. "The Orbital Distribution of Trans-Neptunian Objects beyond 50 Au." *The Astrophysical Journal* 827 (2): L35. https://doi.org/10.3847/2041-8205/827/2/L35.

Pravec, Petr, and Alan W. Harris. 2000. "Fast and Slow Rotation of Asteroids." *Icarus* 148 (1): 12–20. https://doi.org/10.1006/icar.2000.6482.

Richardson, D.C., Z.M. Leinhardt, H.J. Melosh, W.F. Bottke, E. Asphaug. Gravitational Aggregates: Evidence and Evolution. In *Asteroids III*, Eds. Bottke, W., A. Cellino, P. Paolicchi, R.P. Binzel. University of Arizona Press - Tucson, AZ, 501, 515. (2002)

Richardson, D., Elankumaran, P. & Sanderson, R. Numerical experiments with rubble piles: equilibrium shapes and spins. *Icarus* **173**, 349–361 (2005).

Richardson, D. C., Michel, P., Walsh, K. J. & Flynn, K. W. Numerical simulations of asteroids modelled as gravitational aggregates with cohesion. *Planetary and Space Science* **57**, 183–192 (2009).

Rickman, H., S. Marchi, M. F. A'Hearn, C. Barbieri, M. R. El-Maarry, C. Güttler, W.-H. Ip, et al. 2015. "Comet 67P/Churyumov-Gerasimenko: Constraints on Its Origin from OSIRIS Observations." *Astronomy & Astrophysics* 583 (November): A44. https://doi.org/10.1051/0004-6361/201526093.

Rousselot, P. 2008. "174P/Echeclus: A Strange Case of Outburst." *Astronomy & Astrophysics* 480 (2): 543–50. https://doi.org/10.1051/0004-6361:20078150.

Rousselot, P., P. P. Korsun, I. Kulyk, A. Guilbert-Lepoutre, and J.-M. Petit. 2016. "A Long-Term Follow up of 174P/Echeclus." *Monthly Notices of the Royal Astronomical Society* 462 (Suppl 1): S432–42. https://doi.org/10.1093/mnras/stw3054.

Rozitis, B., and S. F. Green. 2013. "The Influence of Global Self-Heating on the Yarkovsky and YORP Effects." *Monthly Notices of the Royal Astronomical Society* 433 (1): 603–21. https://doi.org/10.1093/mnras/stt750.

Samarasinha, Nalin H., and Béatrice E. A. Mueller. 2013. "Relating Changes in Cometary Rotation to Activity: Current Status and Applications to Comet C/2012 S1 (ISON)." *The Astrophysical Journal* 775 (1): L10. https://doi.org/10.1088/2041-8205/775/1/L10.

Sánchez, Paul, and Daniel J. Scheeres. 2014. "The Strength of Regolith and Rubble Pile Asteroids." *Meteoritics & Planetary Science* 49 (5): 788–811. https://doi.org/10.1111/maps.12293.

———. 2016. "Disruption Patterns of Rotating Self-Gravitating Aggregates: A Survey on Angle of Friction and Tensile Strength." *Icarus* 271 (June): 453–71. https://doi.org/10.1016/j.icarus.2016.01.016.

———. 2018. "Rotational Evolution of Self-Gravitating Aggregates with Cores of Variable Strength." *Planetary and Space Science*, April. https://doi.org/10.1016/j.pss.2018.04.001.

———. 2020. "Cohesive Regolith on Fast Rotating Asteroids." *Icarus* 338 (March): 113443. https://doi.org/10.1016/j.icarus.2019.113443.

Sarid, G. ; Volk., K.; Steckloff J.K.; Haris., W.; Womack., M.; Woodney, L.M. (2019) 29P/Schwassmann-Wachmann 1, A Centaur in the Gateway to the Jupiter-family Comets. *ApJL* **883**:L25 (7pp.)

Scheeres, Daniel J. 2007. "The Dynamical Evolution of Uniformly Rotating Asteroids Subject to YORP." *Icarus* 188 (2): 430–50. https://doi.org/10.1016/j.icarus.2006.12.015.

Schwartz, Stephen R., Patrick Michel, Martin Jutzi, Simone Marchi, Yun Zhang, and Derek C. Richardson. 2018. "Catastrophic Disruptions as the Origin of Bilobate Comets." *Nature Astronomy*, March. https://doi.org/10.1038/s41550-018-0395-2.

Di Sisto, R.P.; Brunini, A. (2007) The origin and distribution of the Centaur population. *Icarus* **190**(1), 224 - 235

Statler, T.S. (2009) Extreme sensitivity of the YORP effect to small-scale topography. *Icarus* 202, 502 - 513

Steckloff, Jordan K., and Seth A. Jacobson. 2016. "The Formation of Striae within Cometary Dust Tails by a Sublimation-Driven YORP-like Effect." *Icarus* 264 (January): 160–71. https://doi.org/10.1016/j.icarus.2015.09.021.

Steckloff, Jordan K., Brandon C. Johnson, Timothy Bowling, H. Jay Melosh, David Minton, Carey M. Lisse, and Karl Battams. 2015. "Dynamic Sublimation Pressure and the Catastrophic Breakup of Comet ISON." *Icarus* 258 (September): 430–37. https://doi.org/10.1016/j.icarus.2015.06.032.

Steckloff, Jordan K., and Nalin H. Samarasinha. 2018. "The Sublimative Torques of Jupiter Family Comets and Mass Wasting Events on Their Nuclei." *Icarus* 312 (September): 172–80. https://doi.org/10.1016/j.icarus.2018.04.031.

Steckloff, J.K.; Lisse, C.M.; Safrit, T.K.; Bosh, A.S.; Lyra, W. (2020) The sublimative evolution of (486958) Arrokoth. *Icarus.* In Revision.

Steinberg, Elad, and Re'em Sari. 2011. "Binary YORP Effect and Evolution of Binary Asteroids." *The Astronomical Journal* 141 (2): 55. https://doi.org/10.1088/0004-6256/141/2/55.

Taylor PA, Howell ES, Nolan MC, Thane AA. 2012. The Shape and Spin Distributions of Near-Earth Asteroids Observed with the Arecibo Radar System. AAS/Division for Planetary Sciences Meeting Abstracts 44:302.07

Thomas, Peter C., J. Veverka, Michael J.S. Belton, Alan Hidy, Michael F. A'Hearn, T.L. Farnham, Olivier Groussin, et al. 2007. "The Shape, Topography, and Geology of Tempel 1 from Deep Impact Observations." *Icarus* 187 (1): 4–15. https://doi.org/10.1016/j.icarus.2006.12.013.

Tiscareno, M.S.; Malhotra, R. (2003) The dynamics of Known Centaurs. *AJ* **126**:3122 - 3131

Trigo-Rodríguez, Josep M., D. A. García-Hernández, Albert Sánchez, Juan Lacruz, Björn J. R. Davidsson, Diego Rodríguez, Sensi Pastor, and José A. De Los Reyes. 2010. "Outburst Activity in Comets - II. A Multiband Photometric Monitoring of Comet 29P/Schwassmann-Wachmann 1: Outburst Activity in Comets - II." *Monthly Notices of the Royal Astronomical Society* 409 (4): 1682–90. https://doi.org/10.1111/j.1365-2966.2010.17425.x.

Vincent, J.-B., S. F. Hviid, S. Mottola, E. Kuehrt, F. Preusker, F. Scholten, H. U. Keller, et al. 2017. "Constraints on Cometary Surface Evolution Derived from a Statistical Analysis of 67P's Topography." *Monthly Notices of the Royal Astronomical Society* 469 (Suppl_2): S329–38. https://doi.org/10.1093/mnras/stx1691.

Walsh, K. Rubble Pile Asteroids. *Ann. Rev. of Astronomy and Astrophysics* **56**, 593 - 624 (2018)

Wierzchos, K., M. Womack, and G. Sarid. 2017. "Carbon Monoxide in the Distantly Active Centaur (60558) 174P/Echeclus at 6 Au." *The Astronomical Journal* 153 (5): 230. https://doi.org/10.3847/1538-3881/aa689c.